\documentclass[prl,showpacs,hyperref,superscriptaddress,twocolumn]{revtex4}

\usepackage{color}

\usepackage{amsfonts}
\usepackage{amsmath}
\usepackage{amssymb}
\usepackage{graphicx}
\usepackage{subfigure}
\usepackage{psfrag}
\providecommand{\U}[1]{\protect\rule{.1in}{.1in}}
\providecommand{\U}[1]{\protect\rule{.1in}{.1in}}

\begin{document}

\title{Metastability of Bose and Fermi gases on the upper branch}

\author{Andr\'e LeClair}
\affiliation{
Department of Physics, Cornell University, Ithaca, NY}

\author{Itzhak Roditi}
\affiliation{
Centro Brasileiro de Pesquisas F\'{\i}sicas, Rua Dr. Xavier Sigaud 150, 22290-180 Rio de Janeiro, Rio de Janeiro, Brazil}
\author{Joshua Squires}
\affiliation{
Department of Physics, Cornell University, Ithaca, NY}

\begin{abstract}
\qquad

We study three dimensional Bose and Fermi gases in the upper branch, a phase defined by the absence of bound states in the repulsive interaction regime, within an approximation that considers only two-body interactions. Employing a formalism based on the S-matrix, we derive useful analytic expressions that hold on the upper branch in the weak coupling limit. We determine upper branch phase diagrams for both bosons and fermions with techniques valid for arbitrary positive scattering length.

\end{abstract}

\maketitle

%		DEFINITIONS FOR TEX
%
%%%%%%%%%%%%%%%%%%%%%%%%%%%%%%%%%%%%%%%%%%%%%%%%%%%%%%%%%%%%%%%
%
%
%\def\e{\'e}
%\def\ee{\`e}
%%%%%%%%%%%%%%%%%%%DEFINITIONS%%%%%%%%%%%%%%%%%%%%%%%%%%%%%%%%%
%
\def\oti{{\otimes}}
\def\lb{ \left[ }
\def\rb{ \right]  }
\def\tilde{\widetilde}
\def\bar{\overline}
\def\hat{\widehat}
\def\*{\star}
\def\[{\left[}
\def\]{\right]}
\def\({\left(}		\def\BL{\Bigr(}
\def\){\right)}		\def\BR{\Bigr)}
	\def\BBL{\lb}
	\def\BBR{\rb}
%
%%%%%%%%%%%%%%%%%%%%%%%%%%%%%%%%%%%%%%%%%%%%%%%%%%%%%%%%%%%%%%%
%
\def\zb{{\bar{z} }}
\def\zbar{{\bar{z} }}
\def\frac#1#2{{#1 \over #2}}
\def\inv#1{{1 \over #1}}
\def\half{{1 \over 2}}
\def\d{\partial}
\def\der#1{{\partial \over \partial #1}}
\def\dd#1#2{{\partial #1 \over \partial #2}}
\def\vev#1{\langle #1 \rangle}
\def\ket#1{ | #1 \rangle}
\def\rvac{\hbox{$\vert 0\rangle$}}
\def\lvac{\hbox{$\langle 0 \vert $}}
\def\2pi{\hbox{$2\pi i$}}
\def\e#1{{\rm e}^{^{\textstyle #1}}}
\def\grad#1{\,\nabla\!_{{#1}}\,}
\def\dsl{\raise.15ex\hbox{/}\kern-.57em\partial}
\def\Dsl{\,\raise.15ex\hbox{/}\mkern-.13.5mu D}
%
%%%%%%%%%%%%%%%%%%%%GREEK LETTERS%%%%%%%%%%%%%%%%%%%%%%%%%%%%%%
%
%\def\th{\theta}		\def\Th{\Theta}
\def\ga{\gamma}		\def\Ga{\Gamma}
\def\be{\beta}
\def\al{\alpha}
\def\ep{\epsilon}
\def\vep{\varepsilon}
\def\la{\lambda}	\def\La{\Lambda}
\def\de{\delta}		\def\De{\Delta}
\def\om{\omega}		\def\Om{\Omega}
\def\sig{\sigma}	\def\Sig{\Sigma}
\def\vphi{\varphi}

%
%%%%%%%%%%%%%%%%%%%CALIGRAPHIC LETTERS%%%%%%%%%%%%%%%%%%%%%%%%%
%
\def\CA{{\cal A}}	\def\CB{{\cal B}}	\def\CC{{\cal C}}
\def\CD{{\cal D}}	\def\CE{{\cal E}}	\def\CF{{\cal F}}
\def\CG{{\cal G}}	\def\CH{{\cal H}}	\def\CI{{\cal J}}
\def\CJ{{\cal J}}	\def\CK{{\cal K}}	\def\CL{{\cal L}}
\def\CM{{\cal M}}	\def\CN{{\cal N}}	\def\CO{{\cal O}}
\def\CP{{\cal P}}	\def\CQ{{\cal Q}}	\def\CR{{\cal R}}
\def\CS{{\cal S}}	\def\CT{{\cal T}}	\def\CU{{\cal U}}
\def\CV{{\cal V}}	\def\CW{{\cal W}}	\def\CX{{\cal X}}
\def\CY{{\cal Y}}	\def\CZ{{\cal Z}}

\def\rvac{\hbox{$\vert 0\rangle$}}
\def\lvac{\hbox{$\langle 0 \vert $}}
\def\comm#1#2{ \BBL\ #1\ ,\ #2 \BBR }
\def\2pi{\hbox{$2\pi i$}}
\def\e#1{{\rm e}^{^{\textstyle #1}}}
\def\grad#1{\,\nabla\!_{{#1}}\,}
\def\dsl{\raise.15ex\hbox{/}\kern-.57em\partial}
\def\Dsl{\,\raise.15ex\hbox{/}\mkern-.13.5mu D}
%
%%%%%%%%%%%%%%%%%%%%GREEK LETTERS%%%%%%%%%%%%%%%%%%%%%%%%%%%%%%
%
%%%%%%%%%%%%%%% MATH CHARACTERS %%%%%%%%%%%%%%%%%%%%%%%%%%%%
%
\font\numbers=cmss12
%\font\numbers=cmu10 scaled\magstep1
\font\upright=cmu10 scaled\magstep1
\def\stroke{\vrule height8pt width0.4pt depth-0.1pt}
\def\topfleck{\vrule height8pt width0.5pt depth-5.9pt}
\def\botfleck{\vrule height2pt width0.5pt depth0.1pt}
\def\Zmath{\vcenter{\hbox{\numbers\rlap{\rlap{Z}\kern
0.8pt\topfleck}\kern 2.2pt
                   \rlap Z\kern 6pt\botfleck\kern 1pt}}}
\def\Qmath{\vcenter{\hbox{\upright\rlap{\rlap{Q}\kern
                   3.8pt\stroke}\phantom{Q}}}}
\def\Nmath{\vcenter{\hbox{\upright\rlap{I}\kern 1.7pt N}}}
\def\Cmath{\vcenter{\hbox{\upright\rlap{\rlap{C}\kern
                   3.8pt\stroke}\phantom{C}}}}
\def\Rmath{\vcenter{\hbox{\upright\rlap{I}\kern 1.7pt R}}}
\def\Z{\ifmmode\Zmath\else$\Zmath$\fi}
\def\Q{\ifmmode\Qmath\else$\Qmath$\fi}
\def\N{\ifmmode\Nmath\else$\Nmath$\fi}
\def\C{\ifmmode\Cmath\else$\Cmath$\fi}
\def\R{\ifmmode\Rmath\else$\Rmath$\fi}
%%%%%%%%%%%%%%%%%%%%%%%%%%%%%%%%%%%%%%%%%%%%%%%%%%%%%%%%%%%%%%%%%
 %%%%%%%%%%%%%%%%%% END OF DEFINITIONS %%%%%%%%%%%%%%%%%%%%%%
 %%%%%%%%%%%%%%%%%%%%%%%%%%%%%%%%%%%%%%%%%%%%%%%%%

\def\barray{\begin{eqnarray}}
\def\earray{\end{eqnarray}}
\def\beq{\begin{equation}}
\def\eeq{\end{equation}}

\def\n{\noindent}

\def\Tr{\rm Tr} 
\def\xvec{{\bf x}}
\def\kvec{{\bf k}}
\def\kvecp{{\bf k'}}
\def\omk{\om{\kvec}} 
\def\dk#1{\frac{d\kvec_{#1}}{(2\pi)^d}}
\def\2pid{(2\pi)^d}
\def\ket#1{|#1 \rangle}
\def\bra#1{\langle #1 |}
\def\vol{V}
\def\adag{a^\dagger}
\def\rme{{\rm e}}
\def\Im{{\rm Im}}
\def\pvec{{\bf p}}
\def\fermiS{\CS_F}
\def\cdag{c^\dagger}
\def\adag{a^\dagger}
\def\bdag{b^\dagger}
\def\vvec{{\bf v}}
\def\muhat{{\hat{\mu}}}
\def\vac{|0\rangle}
\def\pcut{{\Lambda_c}}
\def\chidot{\dot{\chi}}
\def\gradvec{\vec{\nabla}}
\def\psitilde{\tilde{\Psi}}
\def\psibar{\bar{\psi}}
\def\psidag{\psi^\dagger} 
\def\m{m_*}
\def\up{\uparrow}
\def\down{\downarrow}
\def\Qo{Q^{0}}
\def\vbar{\bar{v}}
\def\ubar{\bar{u}}
\def\smallhalf{{\textstyle \inv{2}}}
\def\smallsqrt{{\textstyle \inv{\sqrt{2}}}}
\def\rvec{{\bf r}}
\def\avec{{\bf a}}
\def\pivec{{\vec{\pi}}}
\def\svec{\vec{s}} 
\def\phivec{\vec{\phi}}
\def\daggerc{{\dagger_c}}
\def\Gfour{G^{(4)}}
\def\dim#1{\lbrack\!\lbrack #1 \rbrack\! \rbrack }
\def\qhat{{\hat{q}}}
\def\ghat{{\hat{g}}}
\def\nvec{{\vec{n}}}
\def\bull{$\bullet$}
\def\ghato{{\hat{g}_0}}
\def\r{r}
\def\deltaq{\delta_q}
\def\gcharge{g_q}
\def\gspin{g_s}
\def\deltas{\delta_s}
\def\gQC{g_{AF}} 
\def\ghatqc{\ghat_{AF}}
\def\xqc{x_{AF}}
\def\mhat{\hat{m}}
\def\xup{x_2}
\def\xdown{x_1}
\def\sigmavec{\vec{\sigma}}
\def\xopt{x_{\rm opt}}
\def\Lambdac{{\Lambda_c}}
\def\angstrom{{{\scriptstyle \circ} \atop A}     }
\def\AA{\leavevmode\setbox0=\hbox{h}\dimen0=\ht0 \advance\dimen0 by-1ex\rlap{
\raise.67\dimen0\hbox{\char'27}}A}
\def\ratio{\gamma}
\def\Phivec{{\vec{\Phi}}}
\def\singlet{\chi^- \chi^+} 
\def\mhat{{\hat{m}}}

\def\Im{{\rm Im}}
\def\Re{{\rm Re}}

\def\xstar{x_*}

\def\sech{{\rm sech}}

\def\Li{{\rm Li}}

\def\dim#1{{\rm dim}[#1]}

\def\ep{\epsilon}

\def\free{\CF}

\def\Fhat{\digamma}

\def\ftilde{\tilde{f}}

\def\muphys{\mu_{\rm phys}}

\def\xitilde{\tilde{\xi}}

\def\CI{\mathcal{I}}

\def\nhat{\hat{n}}

\def\ef{\epsilon_F}

\def\as{a_s}

\def\diffk{|\kvec - \kvec' |}

\def\dk#1{\frac{d^3 #1}{(2\pi)^3}}
\def\blue#1{{\color{blue}{#1}}}
\def\red#1{{\color{red}{#1}}}

\def\as{a_s}
 
\section{Introduction}

\qquad
It is well known that the two-body interactions of a non-relativistic quantum gas in 3 spatial dimensions can be fully described by the s-wave scattering length, $a_s$. For $\as > 0$, interactions are repulsive and the S-matrix has a pole corresponding
to a bound state or ``molecule".    Thus if one starts with a sample 
consisting of only the fundamental particles,  they will start to combine into
molecules,  which complicates the thermodynamics.    
The ``upper branch" corresponds to $\as > 0$ with the assumption that
the molecules are absent.    This situation has been realized in
experiments \cite{kett2,jochim1,KettJo,strina1,Wild,Makotyn}  and has also been studied theoretically \cite{Duine,Conduit,ohashi1,Pekker,ho2,strina2,ho3,Borzov,Jiang}.  
It is thus natural to inquire under what conditions the upper branch is metastable.

In this paper we will study this question based on 
the formalism developed in \cite{PyeTon}.   
It provides an expression of the free energy at finite temperature and density built on  an integral equation for  the pseudo-energy with a kernel based on the logarithm of the two-body S-matrix at zero temperature.
This integral equation is reminiscent of the Yang-Yang equations used in the Thermodynamical Bethe Ansatz \cite{YangYang}. The formalism is 
  well suited to studying
the upper branch since it is based on the S-matrix and the molecules
are easily eliminated from the thermodynamics by simply not 
including a pseudo-energy for them.

 For both bosons and fermions, the limit $\as \to \pm \infty$ is the so-called unitary limit,  where the theory is scale invariant. The unitary limit has been explored extensively within this formulation of statistical mechanics in \cite{PyeTonUnitary2,Out_unitarity,virial_unitary}, and will therefore not be discussed in this work. Although we will restrict our analysis of the upper branch to $a_s>0$, it is nevertheless important to mention that for bosons the upper branch phase is believed to extend smoothly across unitarity.
 
 In order to determine the boundary between the stable and unstable regions of the upper branch,  we will use the criterion put forward in \cite{ho2,ho3},  namely that the compressibility 
$\kappa$ vanishes. The phase diagram will be determined as a function of
the dimensionless ratios
\beq
\alpha=\frac{\lambda_{T}}{\as}, ~~~~~~~ x  = \frac{\mu}{T}
\eeq  
where  $\lambda_{T}=\sqrt{2 \pi/mT}$ is the de Broglie thermal wave length, $\mu$ the chemical potential, and $\hbar=k_B=1$. 

In the following section we give a brief summary  of the formalism (for further detail see \cite{PyeTon}) and the conventions used in this paper. We then present our results on the stability of the upper branch, and provide an analytic treatment of the integral equation in the weak coupling limit.

\section{Formalism and conventions}

\def\om#1{\omega_#1}
\def\kvec{{\bf k}}
\def\omk{\omega_{\kvec}}
\def\inv#1{\frac{1}{#1}}
\def\d{\partial}

\qquad
In this section we review the main result of \cite{PyeTon}: consistent resummation of two-body scattering leads to an integral equation for a pseudo-energy, whose solution can be used to calculate thermodynamic quantities of interest. We will analyze the upper branch for both bosons and fermions. The Bose gas will be described by the action
\begin{equation}
S=\int d^{3} \mathbf{x} dt \left(i \phi^{\dagger} \partial_{t} \phi-\frac{|{\nabla \phi}|^{2}}{2m}-\frac{g}{2} \left(\phi^{\dagger} \phi \right)^{2} \right),
\end{equation} while for fermions we consider the two-component model defined by:
\begin{equation}
S=\int d^{3} \mathbf{x} dt \left(\sum_{\alpha=\uparrow,\downarrow}i \psi_{\alpha}^{\dagger}\partial_{t} \psi_{\alpha}-\frac{|{\nabla \psi_{\alpha}}|^{2}}{2m}-g \psi_{\uparrow}^{\dagger}\psi_{\uparrow}\psi_{\downarrow}^{\dagger}\psi_{\downarrow} \right).
\end{equation} 
For the non-relativistic theories we will consider,  the two body $S$-matrix can be calculated exactly,  i.e. to all orders in the coupling. Therefore although contributions from many-body ($n>2$) interactions are difficult to calculate, and thus not considered, some non-perturbative aspects are included within this framework.

The filling fractions in this formalism are parametrized in 
terms of a pseudo-energy which has the same form as the free theory. In other words, 
the density can be expressed by: 
\begin{equation}   
 \label{number}
n=\int \frac{d^3 \mathbf{k}} {(2 \pi)^3} \frac{1} {e^{\beta \varepsilon(\mathbf{k})}-s},
\end{equation}
where $\beta=1/T$ is the inverse temperature, s is 1 for bosons and -1 for fermions, and $\varepsilon(\mathbf{k})$ the pseudo-energy. $\varepsilon$ can be interpreted as a self-energy 
correction in the presence of all the particles of the gas that takes into account
multiple scatterings.

Summation of all two-body scattering terms results in an integral equation satisfied by
$\varepsilon$ which we now describe.  
We define the quantity 
\begin{equation} 
  \label{y}
y(\mathbf{k})=e^{-\beta(  \varepsilon(\mathbf{k})  - \omk + \mu)}
\end{equation}
with $\omega_\mathbf{k}=\mathbf{k}^2/2m$, $m$ being the mass of the non-relativistic particles. In terms of $y$, the aforementioned integral equation reads  \begin{equation} \label{inteq}
y(\mathbf{k}) = 1+\beta \int \frac{d^3\mathbf{k}'}{(2\pi)^3}G(\mathbf{k},\mathbf{k}')\frac{z}{e^{\beta \omega_{\mathbf{k}'}}-szy(\mathbf{k}')},
\end{equation} where $z$ is the fugacity. The kernal
\begin{equation}
 \label{kernel}
G(\mathbf{k},\mathbf{k'})=-\frac {16 \pi \sigma} {m. |\mathbf{k}-\mathbf{k}'|} \arctan  \left( \frac {a_	{s}.| \mathbf{k}-\mathbf{k'} |} {2} \right)
\end{equation}
 is derived from the logarithm of the two-body S-matrix:
\beq
\label{Smatrix}
S_{\rm matrix}  ( |\kvec - \kvec' |) =  \frac{ 2/a_s - i |\kvec - \kvec'|}{2/a_s + i |\kvec - \kvec'|}.
\eeq
The factor $\sigma$ in (\ref{kernel}) is $1/2$
 for fermions and $1$ for bosons.

In order to distinguish between the stable and unstable regions of the upper branch, the isothermal compressibility
\beq
\label{comp.1}
\kappa =  - \inv{V} \( \frac{ \d V}{\d p} \)_T  =  - n  \( \frac{ \d n^{-1} }{\d p} \)_T,
\eeq
where $V$ is the volume and $p$ the pressure, will be needed. The second equality above follows since $n=N/V$ with $N$ fixed. The compressibility and particle density can be more conveniently expressed in terms of a scaling function, $q$, of the dimensionless ratios $x$ and $\alpha$:
\begin{gather}
\label{q}
n \,  \lambda_T^3 =  q (x, \alpha)
\\ 
\label{kappa}
\kappa \,  =  \inv{n T}  \frac{ \d_x q }{q}  =
\inv{T} \( \frac{mT}{2\pi} \)^{3/2}   \frac{\d_x q}{q^2}.
\end{gather}
It will also prove useful to define the Fermi surface wavevector $k_F = ( 3 \pi^2 n)^{1/3}$, 
where $n$ is the 2-component density, and the Fermi temperature $T_F =  k_F^2 /2m$ in terms of $q$:
\beq
\label{TFkF}
\frac{T}{T_F}  = \( \frac{4}{3 \sqrt\pi \, q} \)^{2/3}, ~~~~~~~\inv{k_F \as}  = \frac{\lambda_T}{\as}  ( 6 \pi^2 q )^{-1/3}
\eeq
Both of the above expressions also hold for bosons \cite{Out_unitarity}.

Before moving on to a discussion of our results on the upper branch we will put the integral equation and $q$ in more convenient forms. Rotational invariance demands $y$ be a function of  $|\kvec|^2$, thus after rescaling $\mathbf{k}\rightarrow \sqrt{2mT}\mathbf{k}$, the angular integrals in the integral equation (\ref{inteq}) can be performed analytically (see appendix A in \cite{Out_unitarity}).
The result is the following:
\begin{align}
\label{intexpand}
\nonumber 
y(k) = 1 &+ \frac{8}{\pi}   \int_0^\infty  d k' k'    \,  \frac{z}{e^{k'^2} -  s z y (k')}  \\ 
\nonumber 
&\times \Biggl\{   \frac{\alpha}{2 k \sqrt{\pi}} 
\log \[  
\frac{\alpha^2/\pi  +(k  +k' )^2}
{\alpha^2/\pi  + (k -k' )^2}
\] \\ \nonumber 
&- \( \frac{k'}{k} +1 \)  \arctan\(  \frac{\sqrt{\pi}}{\alpha} \( k +k' \) \)  
\\
&- \(  \frac{k'}{k} -1 \)  \arctan\(  \frac{\sqrt{\pi}}{\alpha} \(k - k' \) \)
\Biggr\}.
\end{align}
  Similarly, $q$ can be expressed:
\beq
\label{nhatsc}
q   =  \frac{4}{\sqrt\pi} \int_0^\infty  dk  \, k^2  \, 
  \frac{  y(k) z }{e^{k^2}  -  s\,  y(k) z}. 
\eeq

Finally, note that since the fermion model has two components, in equation (\ref{q}), 
$q \to 2q$ while equations \eqref{TFkF} and (\ref{nhatsc}) remain valid.  Henceforth $q$ will refer to one of the two components.   

\section{Analysis of the upper branch} 
As described in the introduction, the upper branch for Bose and Fermi gases refers to $a_s >0$ with the exclusion of the formation of bound states. In the S-matrix based formalism considered in this paper, removing these states (and the resultant pole in the S-matrix) amounts to simply not including a bound state pseudo-energy. Below we present both boson and fermion phase diagrams for $a_s>0$, as well as an analytic expression for $y$ in the weak coupling limit.

\subsection{A. Phase diagrams}
\begin{figure}[tbp]
\includegraphics[width=0.475\textwidth]{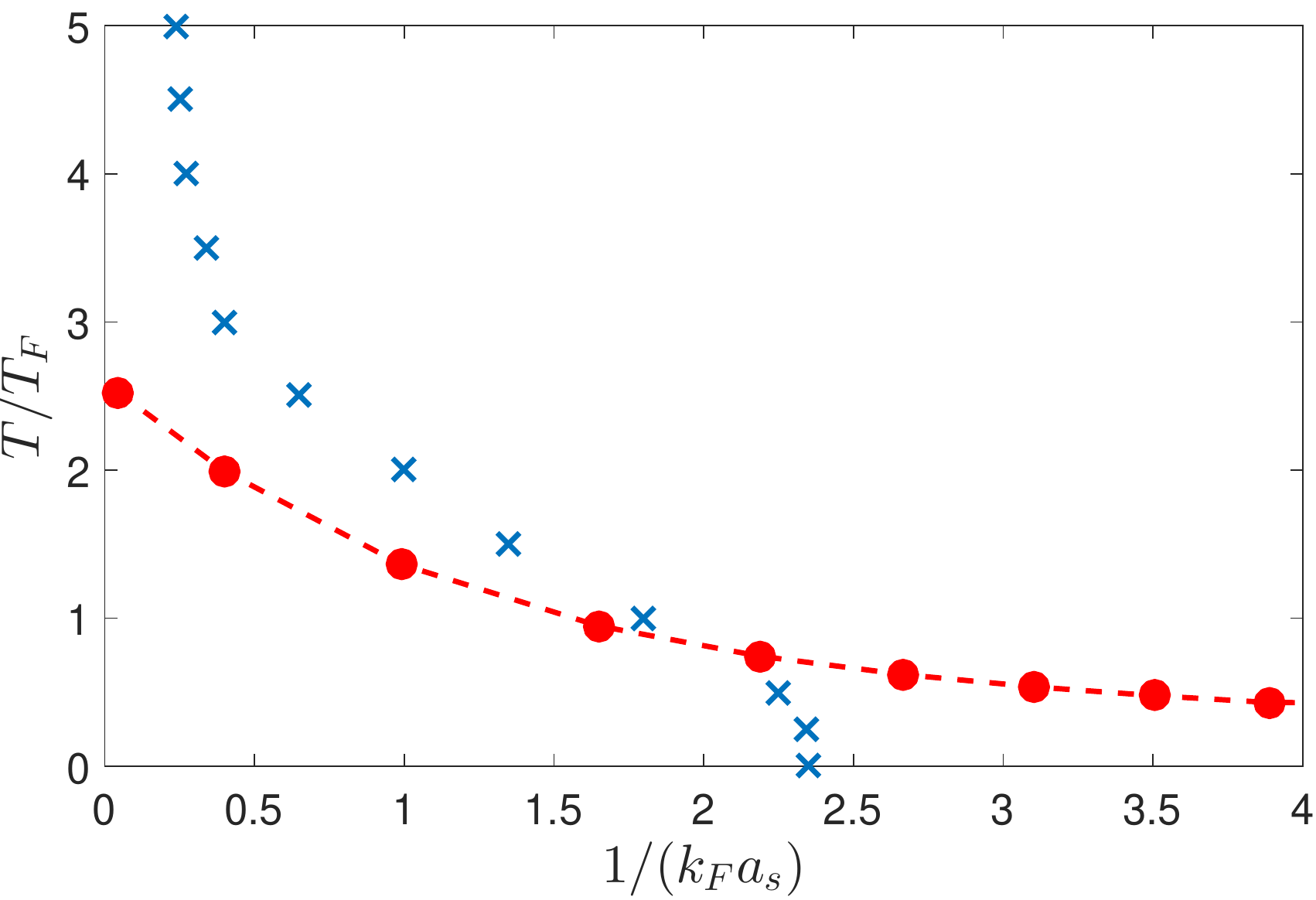}
\caption{Upper branch phase diagram for the Fermi gas. The dashed red curve corresponds to $\kappa = 0$, which defines the boundary between stable and unstable phases. Below this curve the upper branch is unstable. The blue crosses are values of the phase transition estimated from the data presented in \cite{ho2}.}
\label{phasefermi}
\end{figure}

\begin{figure}[tbp]
\centering   
\includegraphics[width=0.475\textwidth]{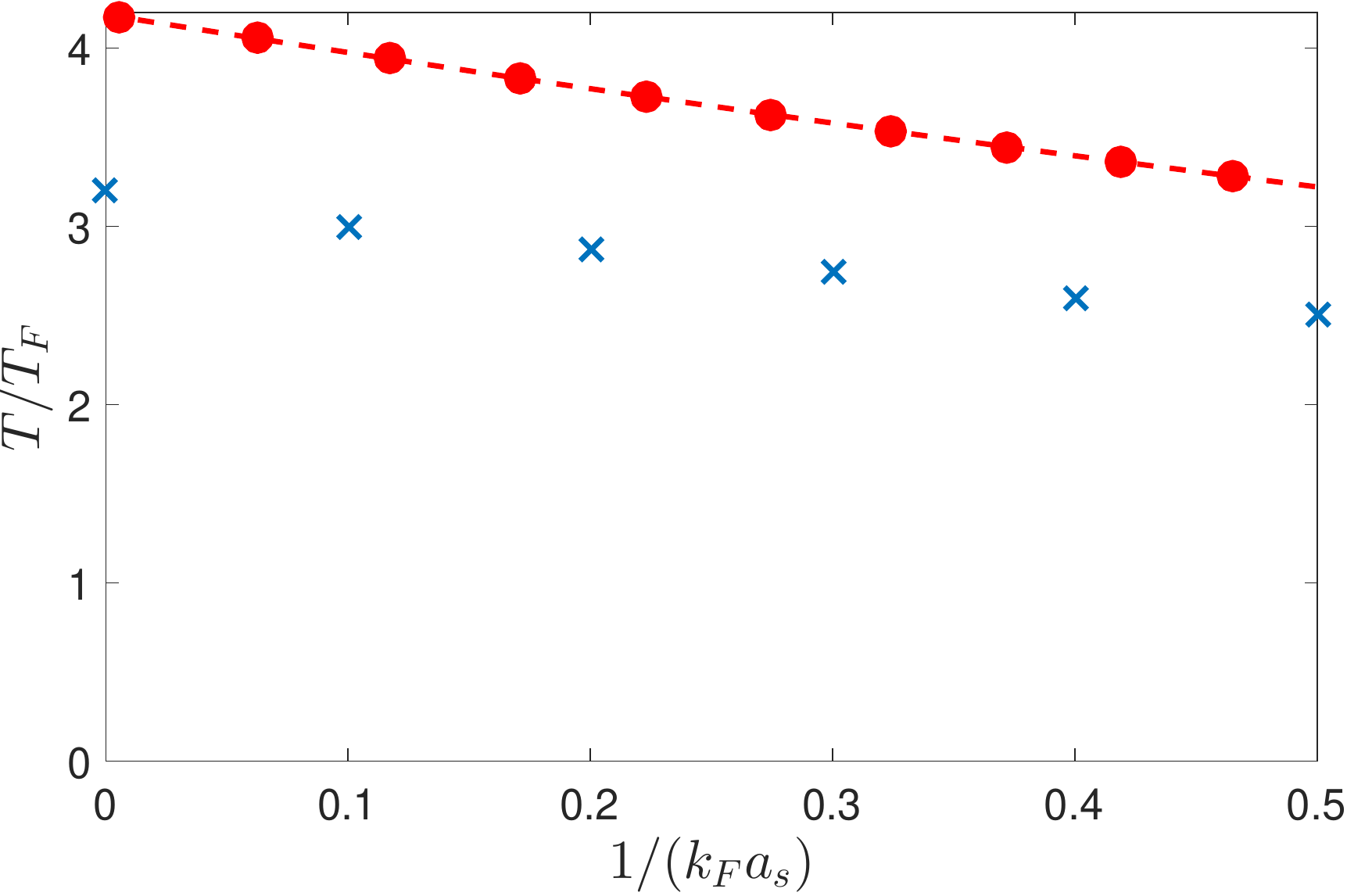}
\caption{Upper branch phase diagram for the Bose gas. The dashed red curve corresponds to $\kappa = 0$, which defines the boundary between stable and unstable phases. Below this curve the upper branch is unstable. The blue crosses are values of the phase transition estimated from the data presented in \cite{ho3}.}
\label{phasebose}
\end{figure}

To determine the metastable region of the upper branch, we have calculated the compressibility as a function of the dimensionless variables $x$ and $\alpha$. This has been achieved by solving the integral equation \eqref{intexpand} and calculating $q$ numerically, from which (\ref{kappa}) is used to determine the stability of the upper branch phase. The curve where the compressibility vanishes provides the boundary between the stable and unstable phases \cite{ho2, ho3}. Our upper branch phase diagram for fermions (bosons) is shown in Figure \ref{phasefermi} (Figure \ref{phasebose}). We emphasize no assumption about the coupling strength has been made in obtaining these phase diagrams.

For fermions we find the phase boundary gradually slopes towards $T/T_F=0$ as $1/k_Fa_s$ increases.  This is consistent with the expectation that the upper branch should be stable
in the limit of zero coupling.     Whereas we find that $T/T_F$ approaches zero asymptotically in the latter limit,   in contrast,  for  the Nozieres-Schmitt-Rink (NSR) based approach employed in \cite{ho2},  it was found that after approximately $1/k_Fa_s=2.5$ the upper branch phase is metastable for all $T/T_F$ (see Figure \ref{phasefermi}).  Though our treatments of the upper branch are quite different, it is encouraging that our results generally agree within the range $0.5<1/k_Fa_s<2.0$. Yet another contrasting result is given in \cite{stoner}, where it's found that the upper branch is always metastable, even at unitarity. 

In order to provide a possible explanation for our weak coupling discrepancy with the excluded molecular pole approximation (EMPA) of Shenoy and Ho, we will consider the limit of very weak coupling, $k_Fa_s \rightarrow 0$. Much of the following analysis is heavily borrowed from section B and appendix A of \cite{stoner}, where it's shown that the EMPA, which begins with the low-fugacity density expansion
\begin{equation}
n_{E}(T,\mu) = n_0(T,\mu)+\partial \Delta P^{(2)}/\partial \mu,
\label{EMPAn}
\end{equation} is identical to an approach which starts with the NSR 2-body interaction contribution to the pressure \begin{equation}
\Delta P^{(2)}=\sum_\mathbf{q} \int_{-\infty}^{\infty}\frac{d\omega}{\pi}\frac{\delta(\mathbf{q},\omega)}{e^{\beta \omega}-1}.
\label{EMPAP}
\end{equation} Note $n_0(T,\mu)$ is the ideal gas density.

The primary obstacle in comparing an NSR-based formalism with our own is that the phase shift $\delta(\mathbf{q},\omega)$ is a complicated function whose definition on the upper branch is not yet agreed upon. In the present limit however, we are concerned only with the leading contribution to the phase shift. Within the EMPA this is simply the vacuum two-body phase shift, $\delta(k) = -\arctan(ka_s)$. Hence after expanding \eqref{EMPAP} in powers of $z^2$, changing variables using the relation $k^2/m = \omega +2\mu-\mathbf{q^2}/4m$, and integrating we obtain
\begin{equation}
\Delta P^{(2)} = \frac{2^{3/2}T}{\lambda_T^3}\sum_{n=1}^{\infty}\frac{z^{2n}}{n^{5/2}}\int_0^{\infty} \frac{dk}{\pi}e^{-\frac{n\beta k^2}{m}}\frac{d\delta (k)}{dk}
\end{equation} which is Eq. (20) in \cite{stoner}. Noting $\frac{d}{dk}\delta (k)\approx -a_s$ as $k_Fa_s \rightarrow 0$ then gives
 \begin{equation}
 \Delta P^{(2)} = -\frac{2T}{\alpha \lambda_T^3}Li_3(z^2).
 \end{equation} Inserting into \eqref{EMPAn} and multiplying through by $\lambda_T^3$ results in the EMPA weak coupling scaling function
 \begin{equation}
 q_{E} = q_0-\frac{4}{\alpha}Li_{2}(z^2).
 \end{equation} 
 
\begin{figure}[tbp]
\includegraphics[width=0.475\textwidth]{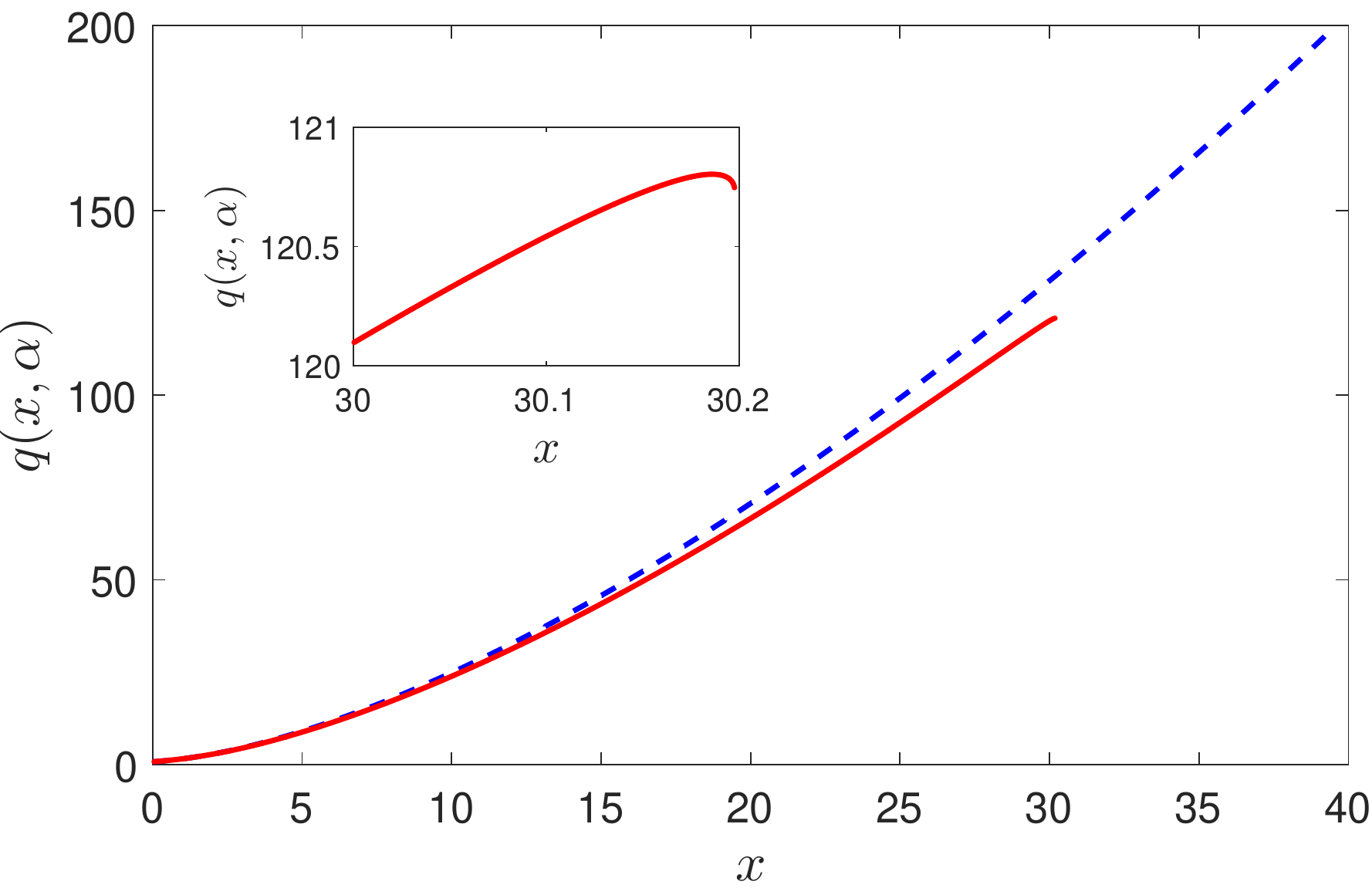}
\caption{Scaling functions vs. $x$ obtained in the $k_Fa_s \rightarrow 0$ limit within the EMPA (dashed blue) and with our formalism (solid red), for $\alpha = 10^3$. The EMPA scaling function is smooth, corresponding to the upper branch being metastable for all $T/T_F$ at weak coupling. Inset: Our $q$ attains a maximum immediately before becoming imaginary. In other words, a bound state is formed and the integral equation no longer has a solution.}
\label{compare}
\end{figure}

 In Figure \ref{compare} we compare $q_E$ and our result in the identical limit, obtained by inserting \eqref{yquadfermi} into \eqref{nhatsc}. $q$ and $q_E$ agree well for small $x$. As $x$ increases, the two results begin to diverge and eventually our $q$ experiences a maximum (where $\kappa = 0$) just before becoming imaginary, signifying an $x$ where the integral equation has no solution. For $\alpha = 10^3$ this maximum occurs around $x \approx 30$, where $T/T_F \approx 0.03$. 
 
 The validity of the EMPA at such large fugacities and small temperatures is unclear, as the NSR approximation is rooted in the virial expansion which relies upon $z \ll 1$. We believe this to be a possible explanation for the differences in our phase diagram and that of \cite{ho2} for fermions at weak coupling: the upper branch phase transition occurs at a very large fugacity, well above the low-fugacity regime where the NSR approximation is most applicable. This also explains why our results are in relative agreement for $1/k_Fa_s \in [0.5,2]$, where the phase transition occurs at much higher temperatures. 

In the bosonic case the $\kappa = 0$ curve is approximately linear, approaching the $T/T_F$ axis near $T/T_F=4.2$. This trend is similar to that calculated in \cite{ho3}, although they are able to extend their results across unitarity, and find the $T/T_F$ intercept to be closer to $T/T_F=3$.

 Both of our phase diagrams take into account only two-body scattering processes. The extent to which many-body interactions alter our findings is presently unknown, but the similarity we see in both the boson and fermion phase diagrams with those of [11,13], which do include many-body effects, suggests the two-body interaction is dominant. For bosons in the unitary limit, the effect of many-body interactions has been estimated to be on the order of a few percent \cite{Borzov}. Unfortunately, experimental and theoretical results alike are limited for the upper branch outside of unitarity.

\subsection{B. Weak coupling limit}
\begin{figure}[tbp]
\centering   
\includegraphics[width=0.475\textwidth]{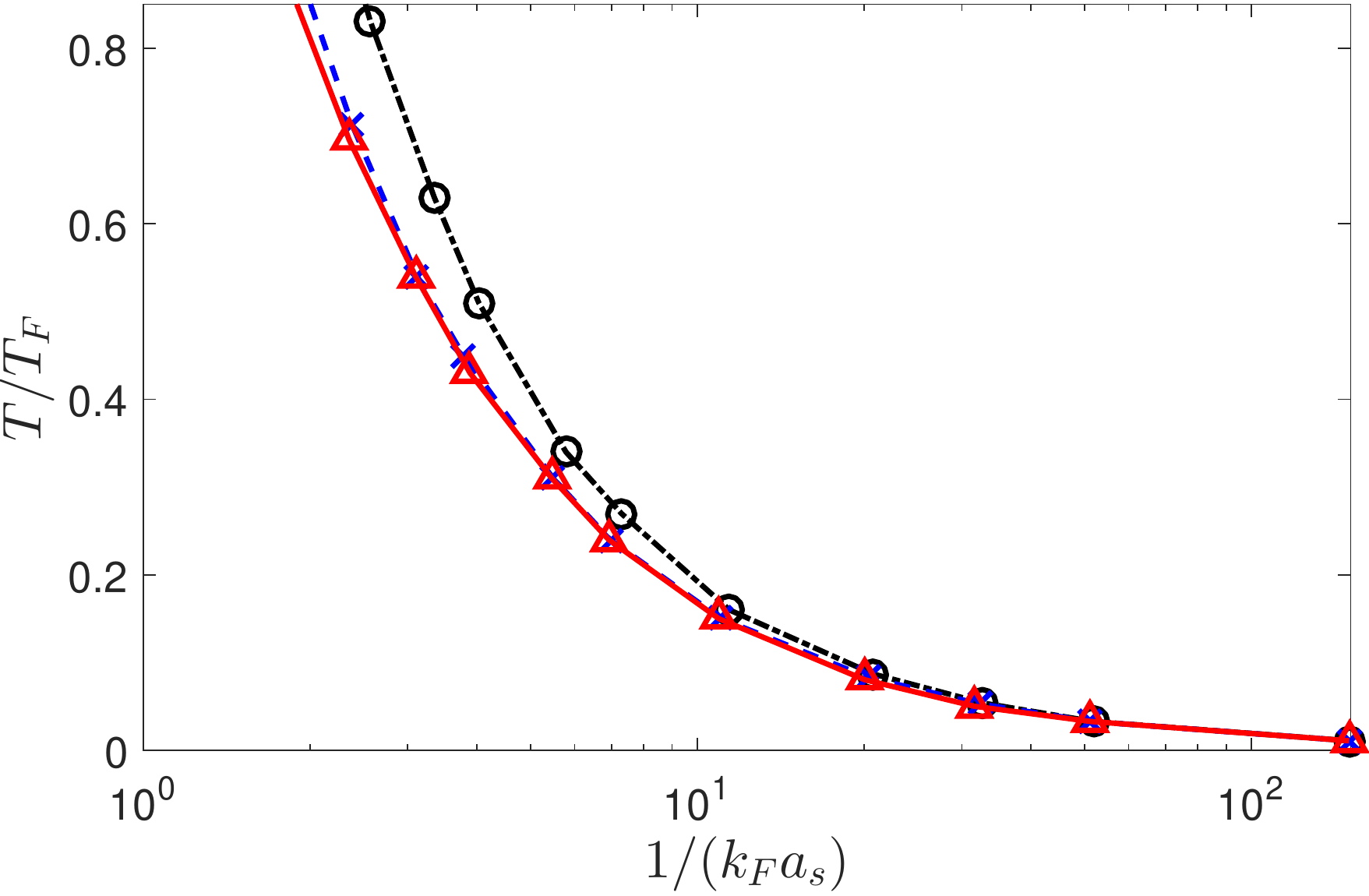}
\caption{Weak coupling behavior of the upper branch phase boundary for fermions. The solid red(dashed blue) curve of zero compressibility was obtained through numerically solving the full integral equation given by \eqref{intexpand}(transcendental equation given by \eqref{ytranscend}). The black dash-dot curve was calculated by using \eqref{cond} to obtain the critical pair $(x_c, \alpha_c)$ corresponding to the phase boundary.}
\label{phasefermiweak}
\end{figure}

For repulsive interactions in the weak coupling regime ($a_s\ll 1$, or equivalently $\alpha \gg 1$) the kernal becomes independent of $k,k'$:
\begin{equation}
G(\mathbf{k},\mathbf{k}') \approx \frac{-8\pi a_s \sigma}{m}. 
\end{equation}
Repeating the manipulations described above in going from (\ref{inteq}) to ({\ref{intexpand}) then gives
\begin{equation*}
y = 1-\frac{16\sigma}{\sqrt{\pi}\alpha s}\int_0^{\infty} dk' \frac{1}{y(k')} \frac{k'^2}{e^{k'^2}/(szy(k'))-1}.
\end{equation*} Since the kernal is a constant, so is $y$ and the remaining momentum integral can be expressed in terms of the polylogarithm:
\begin{equation}
y = 1-\frac{4 \sigma}{\alpha s y} Li_{3/2}(szy).
\label{ytranscend}
\end{equation} Thus we have reduced the integral equation to a transcendental equation, in terms of the scattering length and fugacity, valid for small positive $a_s$. Generally the upper branch phase is stable in the weak coupling limit.

For a free ideal gas $y=1$, as reflected by the form of \eqref{ytranscend} as $\alpha \rightarrow \infty$. Though \eqref{ytranscend} admits no analytic solution for arbitrary $\mu$, an approximate solution can be obtained by setting $y=1$ in the argument of the polylog. Doing so results in a quadratic equation with solutions
\begin{equation}
y=\frac{1}{2}\(1\pm \sqrt{1-\frac{16\sigma Li_{3/2}(sz)}{\alpha s}}\).
\label{yquadratic}
\end{equation}
The positive root must be chosen in order to recover the correct ideal gas behavior.

In the fermionic case \eqref{yquadratic} can be written
\begin{equation}
y=\frac{1}{2}\(1+ \sqrt{\frac{\alpha +8Li_{3/2}(-z)}{\alpha}}\),
\label{yquadfermi}
\end{equation} which suggests an alternate criterion for the stability of the upper branch at weak coupling. For $\alpha \gg 1$ and a given critical $x$ denoted $x_c$, if
\begin{equation}
\alpha = \lvert 8Li_{3/2}(-e^{x_c})\rvert \equiv \alpha_c
\label{cond}
\end{equation} then the pair $(x_c, \alpha_c)$ lies on the phase boundary. If $\alpha<a_c$, then $y$ is complex and the upper branch will be unstable. In Figure \ref{phasefermiweak} the upper branch phase boundary for weakly interacting fermions is computed with this criterion, as well as through the application of \eqref{kappa} with numerical solutions of \eqref{intexpand}} and \eqref{ytranscend}.
All three curves exhibit the same asymptotic behavior as $k_Fa_s$ decreases. At weak coupling, critical points obtained through solving \eqref{ytranscend} are nearly indistinguishable from those calculated by solving the full integral equation, while the condition provided by \eqref{cond} becomes more valid as $\alpha$ increases. 

A similar analysis for bosons is hindered by the fact that the upper branch becomes metastable for all $T/T_F$ before the weak coupling condition can be sufficiently met. Thus for a weakly coupled bose gas, the primary utility of \eqref{ytranscend} and \eqref{yquadratic} lies in computing arbitrary thermodynamic functions, rather than assessing the stability of the upper branch.

\section{Conclusions}
 The formalism developed in \cite{PyeTon} based on the two-body S-matrix, previously applied to quantum gases in the unitary limit and to gases with arbitrary negative scattering length, has been used to study the upper branch. Upper branch phase diagrams for bosons and fermions have been calculated and a simple transcendental equation for the pseudo-energy, valid for repulsive interactions in the weak coupling limit, has been derived. Our findings largely agree with those obtained by other theoretical methods, namely the ``excluded molecular pole approximation" of \cite{ho2,ho3}.
 
  Our methods are most applicable to systems in which two-body interactions dominate. A key open question concerns the degree to which many-body processes affect the metastability of the upper branch \cite{Borzov,Jiang}. We believe the results obtained in this work will be useful in guiding future experiments on the upper branch, both in answering this question and others.
\section{Acknowledgments}
The authors acknowledge financial support from
CNPq, CAPES, and FAPERJ. This work is partly funded
by a Science Without Borders-CNPq grant.
%---------------------------------------------------------------------------------------------------------

\end{document}